\documentclass[a4paper,11pt]{article}
\pdfoutput=1 

\usepackage{jinstpub} 

\hypersetup{
    colorlinks,
    linkcolor={red!50!black},
    citecolor={blue!50!black},
    urlcolor={blue!80!black}
}
\usepackage{graphicx}
\usepackage{url}
\usepackage{orcidlink}

\title{ Analysis of Atomic Charge State and Atomic Number for VAMOS++ Magnetic Spectrometer using Deep Neural Networks and Fractionally Labelled Events}

 \author{M.~Rejmund\,\orcidlink{0009-0009-8626-8756}}
\author{A.~Lemasson\,\orcidlink{0000-0002-9434-8520}} 
 \affiliation{GANIL, CEA/DRF - CNRS/IN2P3, Bd Henri Becquerel, BP 55027, F-14076 Caen Cedex 5, France}

\emailAdd{mrejmund@ganil.fr}

\abstract{
The VAMOS++ magnetic spectrometer is a multi-parametric system that integrates ion optical magnetic elements 
with a multi-detector stack. The magnetic elements, along with the tracking and timing detectors and 
the trajectory reconstruction method, provide the analysis of the magnetic rigidity, the trajectory 
length between the beam interaction point and the focal plane of the spectrometer, and the related 
velocity and mass-over-charge ratio. The segmented ionization chamber provides the energy measurements 
necessary to analyze the atomic charge state and atomic number. 
However, this analysis critically suffers from inherent limitations due to the variable thickness and 
non-uniformity of the entrance window  of the ionization chamber and other detector imperfections. 
Conventionally, this meticulous, detailed analysis is exceptionally tedious, often requiring several months 
to complete.
We present a novel method utilizing deep neural networks, trained on an experimental dataset with 
only a small fraction of precisely labeled events for the lowest and best-resolved atomic charge states 
or numbers. This innovative approach enables the networks to autonomously and accurately classify 
the remaining events.
This method drastically accelerates the acquisition of high-resolution atomic charge state and atomic number
spectra, reducing analysis time from months to mere hours. Crucially, by discarding human bias, this approach 
ensures standardized, optimal, and reproducible results with unprecedented efficiency.
}


\begin{document}
\maketitle
\flushbottom

\section{Introduction}
\label{sec:intro}

Magnetic spectrometers are indispensable tools in nuclear structure and reaction mechanism studies, enabling 
the precise identification and characterization of nuclei produced in nuclear collisions. They provide 
event-by-event isotopic identification of reaction products, crucially ensuring clean selection for subsequent analysis.
The spectrometers like VAMOS++~\cite{Pullanhiotan2008, Rejmund2011} at 
GANIL and  PRISMA~\cite{Montagnoli2005} at LNL Legnaro
are particularly suited for reactions occurring near the Coulomb barrier, where a large angular and momentum acceptance 
is necessary for efficient collection of the nuclear reaction products of interest. 
However, this large angular and momentum acceptance is associated with the need for large optical elements 
in the spectrometer, which often results in a design that exhibits highly non-linear optics. To overcome this challenge, 
ray-tracing techniques are employed to reconstruct the trajectories of the detected heavy ions~\cite{Lemasson2023a, Rejmund2025a}. 
This process allows for the determination of the magnetic rigidity and related observables for these ions.

The VAMOS++ spectrometer, with its large angular and momentum acceptance, is specifically designed to provide 
isotopic identification of incoming ions, resolving them by atomic mass number ($A$), atomic charge 
state ($q$), and atomic number ($Z$).
The magnetic rigidity ($B\rho$), the trajectory length ($l$) between the beam interaction point and the focal plane 
of the spectrometer,  the ion’s velocity ($v$), and the related mass-over-charge ratio ($A/q$),
are obtained using the tracking and timing detectors and the trajectory reconstruction 
methods~\cite{Lemasson2023a, Rejmund2025a}.  
However, achieving accurate isotopic identification critically depends on the precise analysis of the atomic 
charge state ($q$) and atomic number ($Z$), both derived from energy measurements within the segmented 
ionization chamber~\cite{Lemasson2023a}. This work introduces a groundbreaking method to significantly 
simplify, accelerate, and enhance the reliability of this crucial analysis step.

\section{Motivation}
\label{sec:moti}

\begin{figure}[b]\centering
\includegraphics[width=0.6\textwidth]{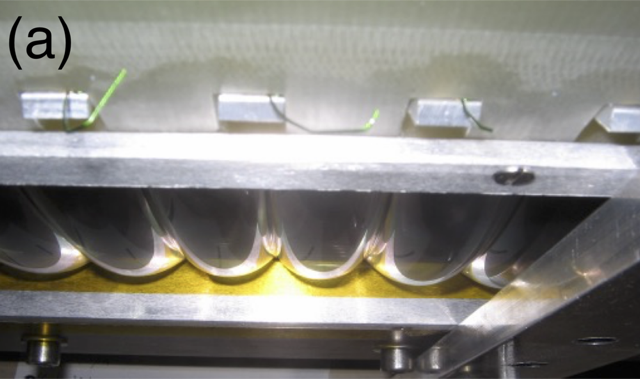}
\hspace{1cm}
\includegraphics[width=0.29\textwidth]{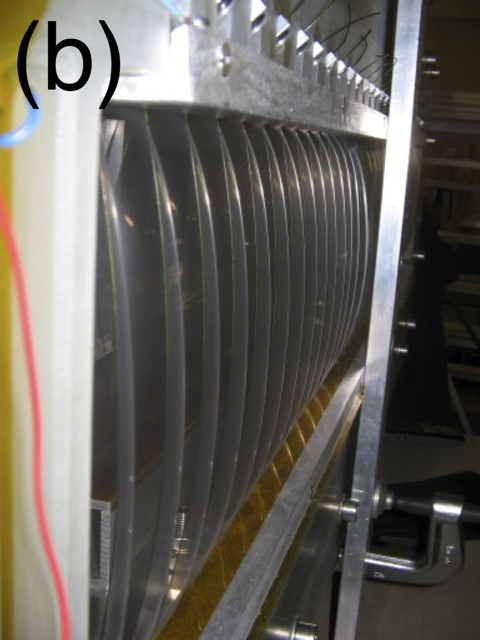}
\caption{Deformation of the ionization chamber windows:
This figure presents the vertical (a) and horizontal (b) views of the ionization chamber window after a test conducted 
on the test bench.
\label{fig:ICWind}}
\end{figure}

VAMOS++ operates with heavy ions ranging from carbon ($Z=6$) to uranium ($Z=92$) at energies between 
approximately $2$~MeV/u and $10$~MeV/u. However, achieving accurate energy measurements from the 
segmented ionization chamber data~\cite{Lemasson2023a} is severely complicated by the highly variable 
thickness and non-uniformity of the material preceding the active volume. This is particularly challenging 
due to the unpredictable shape and amplitude of large-area window deformations, which are highly susceptible 
to pressure changes. To maximize the number of identifiable ions in terms of atomic number by utilizing 
the correlations between energy loss and residual energy, the thickness of the windows is minimized to reduce 
the population of the unresolved low-energy phase space associated with the Bragg peak. 
The VAMOS++ focal plane detectors cover an active area of $1000 \times 150~\text{mm}^2$. Within this large area, 
the deformation of the ionization chamber windows alone can introduce a significant variation of approximately 
$3~\%$ in the measured total energy. Figure~\ref{fig:ICWind} vividly illustrates these large amplitude deformations 
in both vertical (a) and horizontal (b) views, highlighting the extent of this challenge.
Additionally, any imperfections of the detector, such as charge collection or non-linearities in the electronics, must 
also be considered during the energy measurement analysis process. 

Achieving a precise absolute energy calibration of the ionization chamber before each experiment is impractical. 
This is due to the unavailability of radioactive sources covering the required heavy ion and energy ranges. Moreover 
detailed pre-experiment mapping of the  active region of the ionization chamber via elastic scattering of specific ions is prohibitively costly and time-consuming. Therefore, our analysis strategically leverages the inherent self-consistency 
of each experiment's data, augmented by fundamental physics principles and approximate online calibration coefficients 
for raw energies from each ionization chamber row.

\section{Experimental methods}
\label{sec:Exp}

\begin{figure}[b]\centering
\includegraphics[width=0.49\textwidth]{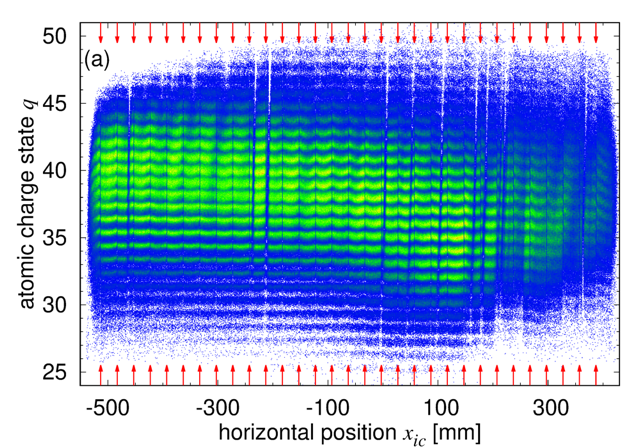}
\includegraphics[width=0.49\textwidth]{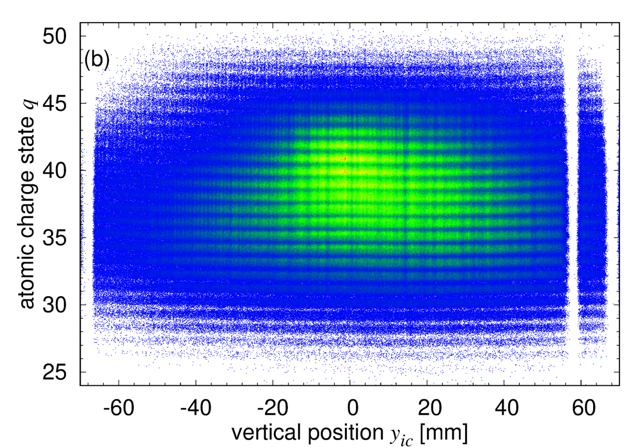}
\caption{Non-uniformity of the material:
Two-dimensional correlation plot between atomic charge state ($q$) and 
(a) the horizontal position ($x_{ic}$) and (b) vertical position ($y_{ic}$) on the entrance window 
of the ionization chamber.
The red arrows in panel (a) indicate the positions of the vertical window holding wires, 
spaced by $30$~mm,  shown in Figure~\ref{fig:ICWind}.
The charts (a) and (b) were obtained using approximate calibration coefficients ($C_i$) for the raw energies 
($E_{ic_i}$) measured by each row of the ionization chamber.
\label{fig:QXICYIC}}
\end{figure}

\begin{figure}[b]\centering
\includegraphics[width=0.49\textwidth]{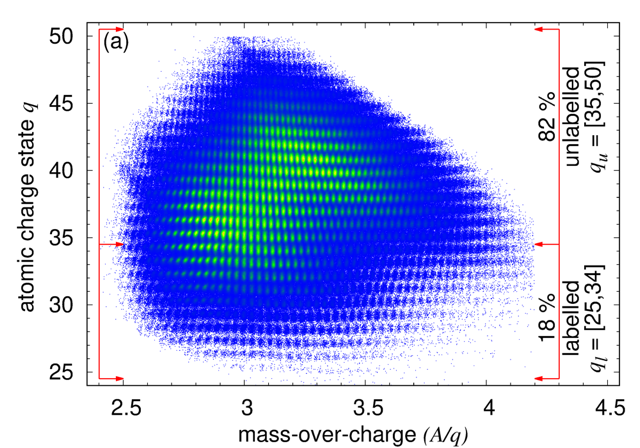}
\includegraphics[width=0.49\textwidth]{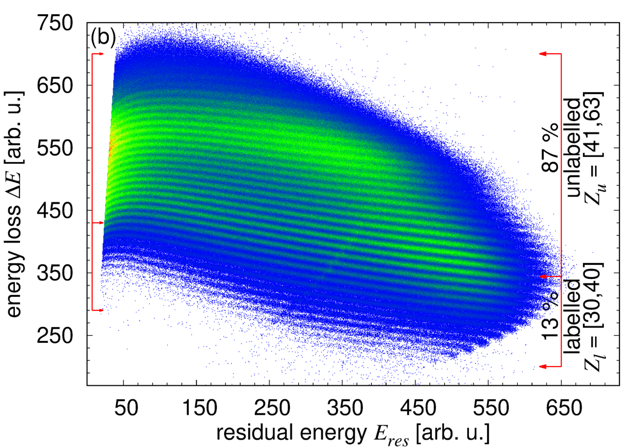}
\caption{Initial identification charts:
(a) Two-dimensional correlation plot between atomic charge state ($q$) and mass-over-charge, ($A/q$). 
Each accumulation for a given atomic charge state corresponds to a specific atomic mass number. 
(b) Two-dimensional energy correlation plot between the energy loss ($\Delta E$) and residual energy ($E_{res}$)
resulting in correlated bands for each atomic number.  For the lowest $E_{res}$, the unresolved part of the spectra,
associated with the Bragg peak, was omitted.
The charts (a) and (b) were obtained using approximate calibration coefficients ($C_i$) for the raw energies 
($E_{ic_i}$) measured by each row of the ionization chamber.
\label{fig:QMQdEE}}
\end{figure}

The experimental data employed in this study was acquired from during the E826 GANIL experiment~\cite{DataE826}
designed to detect and identify the fission fragments generated in fusion-fission and transfer-fission reactions. 
These reactions were induced by a $^{238}$U beam with an energy of $5.88$~MeV/u on a $0.5$~mg/cm$^2$ thick 
$^9$Be target. The VAMOS++ spectrometer was positioned at an angle of $20^\circ$ relative to the beam axis.
The fission fragments with atomic numbers spanning approximately 30 to 60 were detected by the VAMOS++ spectrometer, 
with energies ranging from approximately 4~MeV/u to 10~MeV/u. This resulted in a relatively broad distribution 
of atomic charge states from $25$ to $50$. The ion tracking and timing were ensured by the dual position-sensitive 
Multi-Wire Proportional Chamber (MWPC) telescope~\cite{Vandebrouck2016}, positioned at the entrance of VAMOS++ and 
two large-area position-sensitive MWPCs positioned in the focal plane. The ionization chamber consisted of ten rows with thicknesses as follows:  $4 \times 30$~mm, $2\times 60$~mm,  $3\times 120$~mm,  and $100$~mm, respectively. 
It was operated at the pressure of $100$~mbar of  tetrafluoromethane ($\text{CF}_4$) gas with a $2.5~\mu$m thick 
Mylar window.

Building upon recent advancements, we utilized a novel deep neural network-based method for trajectory reconstruction~\cite{Rejmund2025a}. This method, trained on a theoretical dataset from standard magnetic 
ray-tracing code, provides universal trajectory reconstruction. We employed this approach to precisely determine 
the magnetic rigidity ($B\rho$), trajectory length ($l$) between the beam interaction point and the spectrometer's 
focal plane, ion velocity ($v$), and the crucial mass-over-charge ratio ($A/q$).

During the standard analysis process, the energy loss ($\Delta E$) and residual energy ($E_{res}$), 
are obtained as follows:
\begin{equation} 
\Delta E = \sum_{i=0}^{4} C_i \cdot E_{ic_i}
\label{dE}
\end{equation} 
 and
\begin{equation} 
 E_{res} = \sum_{i=5}^{9} C_i \cdot E_{ic_i}, 
\label{E}
\end{equation}
where: $E_{ic_i}$ are the raw energies measured by the ionization chamber rows and
$C_i$ are the calibration coefficients.
While we assume a linear calibration for simplicity, it's important to note that a truly accurate calibration would 
ideally account for any inherent non-linearities.
The atomic number ($Z$) is derived from the energy correlation between the measured energy loss ($\Delta E$) and 
residual energy ($E_{res}$).
The total energy ($E_{tot}$) in accordance with the measured magnetic rigidity ($B\rho$) and velocity ($v$) contains 
the unmeasured energy ($E_{um}$) lost in the matter in front of the active volume of the ionization chamber 
and is expressed as:
\begin{equation} 
 E_{tot} = E_{um} + \Delta E  + E_{res}.
\label{Etot}
\end{equation} 
 The atomic charge state ($q$) is obtained from the equation:
\begin{equation} 
q = \frac{E_{tot}}{1~\text{u} \cdot (\gamma-1) \cdot (A/q)}, 
\label{qEq}
\end{equation} 
where: $\text{u} = 931.494$~MeV/$c^2$ is the unified atomic unit and $\gamma$ is the Lorentz factor.

The approximate energy calibration is relatively straightforward to obtain and is typically performed 
during the online analysis of the experimental data. 
The two-dimensional correlation spectra between the obtained atomic charge state ($q$) and the horizontal 
($x_{ic}$) and vertical ($y_{ic}$) positions on the entrance window of the ionization chamber are presented in Figure~\ref{fig:QXICYIC}(a) and (b), respectively. In panel (a), the position of the vertical window holding wires ($30$~mm spacing) is indicated by the red arrows. The impact of the deformation of the entrance window of 
the ionization chamber correlated with the positions of the holding wires on ($q$)  is clearly visible in 
Figure~\ref{fig:QXICYIC}(a).  Similarly, the parabolic shape of the ($q$) lines in panel (b) of Figure~\ref{fig:QXICYIC} 
corresponds to the window deformation in the vertical direction.
The corresponding identification charts are presented in 
Figure~\ref{fig:QMQdEE}. The correlation plot between atomic charge state ($q$) and mass-over-charge, ($A/q$) is 
depicted in Figure~\ref{fig:QMQdEE}(a), where each accumulation for a  given $q$ corresponds to a specific atomic 
mass number. The energy correlation between the energy loss ($\Delta E$) and residual energy ($E_{res}$) is presented in 
Figure~\ref{fig:QMQdEE}(b), where the correlated bands correspond to specific atomic numbers.
The identification of distinct nuclei can typically be verified by employing the coincident characteristic $\gamma$ rays.

Achieving the highest possible resolution for atomic number ($Z$) and atomic charge state ($q$) necessitates 
meticulous analysis of energy measurements, which in turn enables the cleanest isotopic selection of detected ions. 
Traditional methods involve iterative corrections, meticulously uncovering dependencies of $Z$ and $q$ on various 
parameters like horizontal/vertical positions and mass-over-charge, to compensate for detector and calibration deficiencies. However, this multivariate correlation analysis is exceedingly time-consuming, often spanning several months, and inherently susceptible to human bias.

\section{Application of deep neural networks}
\label{sec:AppNN}

To overcome these significant challenges, this work introduces a novel method for analyzing atomic charge state and 
atomic number using advanced deep neural networks. These networks, inspired by biological neurons, excel at discerning 
complex, multidimensional patterns. Crucially, they can not only learn and replicate provided patterns but also extrapolate 
effectively to previously unseen data sequences.

Various neural network-based approaches are applicable for particle identification tasks. Among supervised learning methods 
that utilize conventional, purely data-driven neural networks i.e., without explicit incorporation of prior physical laws pertinent to 
the data, the two most prevalent model forms are classification and regression~\cite{kinsley2020, Subasi2020}. 
Classification models are employed when the objective 
is to assign an input to one of several discrete classes; their output typically represents a probability distribution over these classes. 
In contrast, regression models are used when the goal is to predict a continuous numerical value. Both classification and regression 
models can be trained on either experimental or simulated data, depending on the specific application. Typically, all events used 
in the training procedure for these supervised methods are fully labeled.

In this work, a regression neural network model was specifically employed. The primary objective of this model was to provide 
the atomic charge state and atomic number on an event-by-event basis while simultaneously minimizing the widths of their 
corresponding distributions. 
This optimization directly leads to a significant enhancement in the separation quality of these distinct states. 
A unique and distinguishing aspect of our method is the fractional labeling of the events used to train the neural network model. 
This approach allowed the model to be predominantly trained autonomously in an auto-supervised mode, thereby reducing the reliance 
on fully pre-labeled datasets.

We have used a general dense deep feed-forward neural network architecture~\cite{kinsley2020, Subasi2020} 
of the form $N_l \times N_u$, comprising $N_l $ layers and $N_u$ units (neurons) per layer, followed 
by a single output unit, similarly as described in Ref.~\cite{Rejmund2025a}. For the analysis of the atomic charge state ($q$)
and atomic number ($Z$) the architecture was chosen as  $N_l \times N_u = 8 \times 32$. 
The networks were trained on an experimental dataset. The input to the deep neural network consisted of variables 
(details provided in Sections~\ref{sec:AnQ} and \ref{sec:AnZ}) such as raw energies measured by ionization chamber rows 
or horizontal and vertical positions. No relationships such as those given in equations~\ref{dE}, \ref{E}, or \ref{qEq}
nor calibration coefficients were provided to the network; thus, the correlations such as $q-(A/q)$ or $\Delta E-E_{res}$ 
discussed above were not known by the network.

A key innovation of our approach is the minimal labeling strategy. Only a small fraction of events, specifically those 
corresponding to the lowest and best-resolved atomic charge states or numbers, were precisely labeled with their 
integer values. This drastically minimizes potential bias from erroneous labeling. These labeled events serve as a minimal 
guiding principle for the network; the vast majority of events were then processed autonomously during training. 
This choice stems from the understanding that deep neural networks can readily perceive the fundamental properties 
of the experimental system from limited, high-quality labeled data. This perception allows the networks to effectively 
organize the remaining unlabeled events, recognizing their inherent integer nature.
The labeled and unlabeled events were provided to the networks 
simultaneously, following the experimental repetition frequency. The dataset used to train the neural networks
consisted of $5\times 10^7$ events. The  dataset was randomly partitioned into the training set (80~\%) and 
the validation set (20~\%).

\subsection{Analysis of atomic charge state}
\label{sec:AnQ}

\begin{figure}[b]\centering
\includegraphics[width=0.49\textwidth]{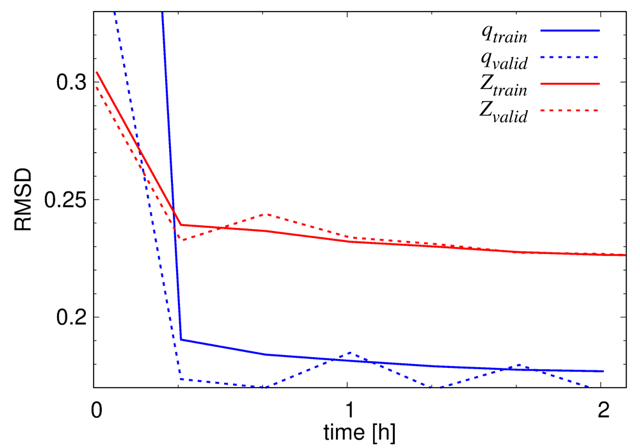}
\caption{Training Convergence: 
The root mean square deviation (RMSD) values for $q$ and $Z$ are plotted as a function of time.
The index $train$ denotes the training curve, while $valid$ denotes the validation curve.
\label{fig:Conv}}
\end{figure}

To analyze the atomic charge state ($q$) one starts with the approximate calibration coefficients ($C_i$)
 (equations~\ref{dE} and \ref{E}).
The resulting correlation between the atomic charge ($q$) (equation~\ref{qEq}) and the mass-over-charge ($A/q$) 
is depicted in Figure~\ref{fig:QMQdEE}(a). 
It is evident that the atomic charge states are poorly resolved and exhibiting clear dependencies on the mass-to-charge ratio.
The correlation between $q$ and $(A/q)$ provides a clear basis for labeling the lowest, most distinct values of $q$, 
assigning them as integer target values for neural network training. This strategy leverages the fact that lower $q$ values 
are generally better resolved than higher ones.
In our case, we have labeled the atomic charge states within the range $q_l = [25,34]$, which corresponds to 
the repetition probability of approximately $18~\%$. The remaining portion of the data in the range $q_u = [35,50]$ 
remained unlabeled. This selection is depicted in Figure~\ref{fig:QMQdEE}(a) using red arrows. Depending on the resolution 
obtained from the initial calibration, the selection gate for each chosen $q_l$ can be either wide or narrow, centered at the locus 
of the centroids. The omitted events can remain unlabeled.

To evaluate the gradient descent, for the labeled events in the range $q_l=[25,34]$, the training target 
was the corresponding $q_t=q_l$, while for the unlabeled events in the range $q_u=[35,50]$, it was 
the nearest integer value to the predicted in each learning step $q_p$, $q_t=\lfloor q_{p}+0.5 \rfloor$.
The input to the deep neural network comprised the following $17$ variables:
\begin{itemize}
\item the raw energies obtained by the ionization chamber ($E_{ic_i}$) with $i=0 \dots 9$,
\item the reciprocal of the mass-over-charge, ($A/q$), and of
the Lorenz term, $(\gamma - 1)$, 
\item the horizontal and vertical position on the entrance window of the ionization chamber, 
($x^{cnt}_{ic} = d \cdot \lfloor  x_{ic}/d +0.5 \rfloor$,  $\delta x_{ic} = x_{ic} - x^{cnt}_{ic}$)
 and ($y_{ic}$),
position information in the focal plane of VAMOS++, where $d =30$~mm is the spacing between 
the window holding wires,
\item  the horizontal and vertical position on the exit window of the dual position-sensitive MWPC telescope~\cite{Vandebrouck2016},
($x_{mw}$) and ($y_{mw}$), positioned at the entrance of VAMOS++.
\end{itemize}
The horizontal position on the entrance window of the ionization chamber was expressed as the distance
to the closest center point between two holding wires ($\delta x_{ic}$) and the position of the 
corresponding center point ($x^{cnt}_{ic}$) .
The inclusion of horizontal and vertical positions explicitly enables the neural networks to account for and correct 
the non-uniformity of the matter layers. The output of the neural networks provides the atomic charge state, $q_{NN}$.
The training convergence of the neural network was evaluated in terms of the root mean square deviation (RMSD).

\begin{figure}[t]\centering
\includegraphics[width=0.49\textwidth]{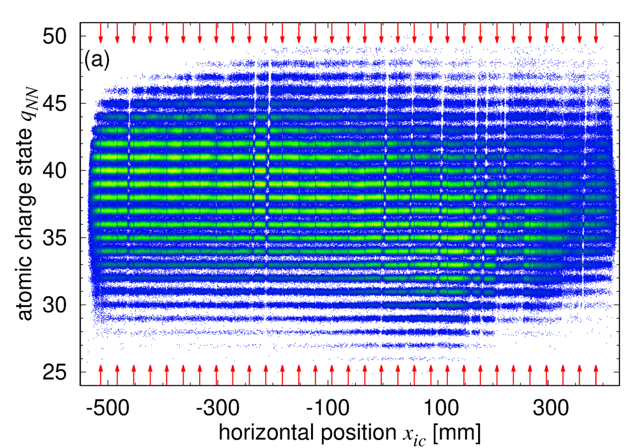}
\includegraphics[width=0.49\textwidth]{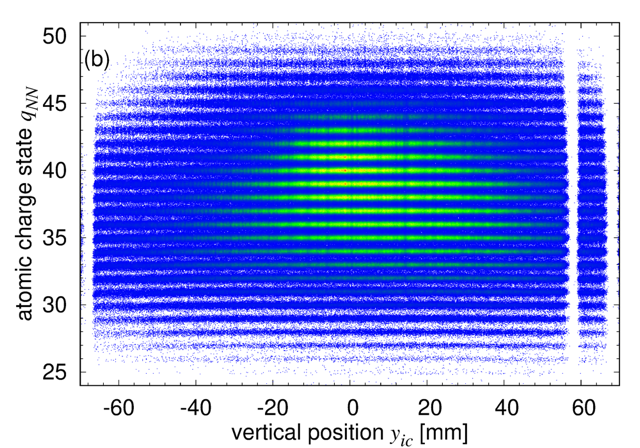}
\includegraphics[width=0.49\textwidth]{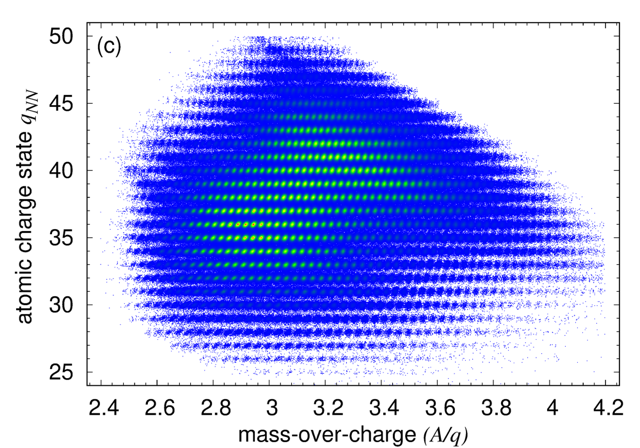} 
\includegraphics[width=0.49\textwidth]{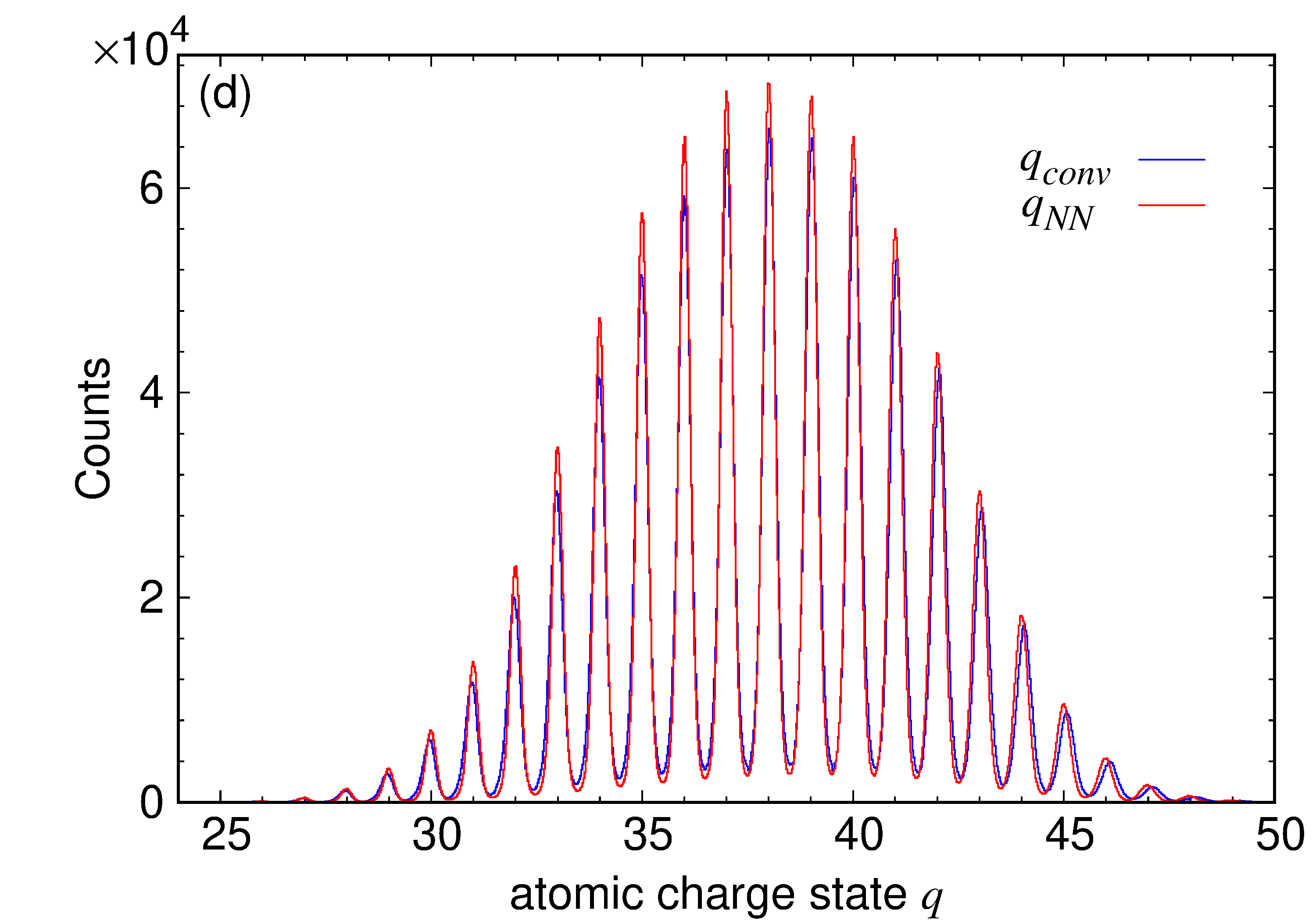}
\caption{Results of the neural network analysis of the atomic charge state:
(a) and (b) The same as in Figure~\ref{fig:QXICYIC}(a) and (b) but for $q_{NN}$ instead of $q$,
(c) The same as in Figure~\ref{fig:QMQdEE}(a) but for $q_{NN}$ instead of $q$, 
(d) One-dimensional spectrum of atomic charge state, in red $q_{NN}$ and in blue $q_{conv}$ obtained using the 
conventional analysis method.
\label{fig:QMQNN}}
\end{figure}

The duration of one training epoch was about $20$~s. The training and validation RMSD was evaluated every 60 epochs ($20$~min)
with early stopping with a tolerance $0.001$ and a patience of $60$ epochs.
The training convergence curve of the RMSD for $q$ as a function of time is depicted in Figure~\ref{fig:Conv}, where the 
index $train$ denotes the training and $valid$ denotes the validation.
Remarkably, the neural network rapidly discerns the primary system characteristics within approximately $20$ minutes. 
Following this initial rapid learning phase, convergence proceeds at a significantly slower pace, indicating a detailed 
refinement process. The validation curve closely mirrors the training curve, demonstrating robust generalization.
During the training, the neural network achieved convergence, resulting in an RMSD for $q$ of $0.177$.

The correlations between the atomic charge state ($q_{NN}$) obtained by the neural network and the horizontal
and vertical position on the entrance of the ionization chamber are shown in Figures~\ref{fig:QMQNN}(a) and \ref{fig:QMQNN}(b), respectively.
Figure~\ref{fig:QMQNN} clearly demonstrates a dramatic improvement in the definition of the atomic charge state. 
The widths of $q$ for individual atomic mass numbers are significantly narrower and  the detrimental dependencies on
the positions, prominently observed in Figure~\ref{fig:QXICYIC}, are completely discarded.
The correlation between the atomic charge state ($q_{NN}$)  and the mass-over-charge
($A/q$) is depicted in Figure~\ref{fig:QMQNN}(c). 
Figure~\ref{fig:QMQNN}(c) also demonstrates an improvement in the definition of the atomic charge state. 
The detrimental dependencies on the mass-to-charge ratio, observed in Figure~\ref{fig:QMQdEE}(a), 
are also discarded.
The corresponding one-dimensional spectrum of $q_{NN}$ is presented in Figure~\ref{fig:QMQNN}(d)
and compared to $q_{conv}$ obtained in a complete process of the conventional analysis method.
The resulting resolution in terms of $\text{FWHM}(q_{NN})/q_{NN}$ ranges from $1.0~\%$, for the highest atomic charge state, 
to $1.2~\%$, for the lowest atomic charge state, respectively. The overall improvement in terms of 
$\text{FWHM}$ for $q_{NN}$ with respect to $q_{conv}$ is $9~\%$.

\subsection{Analysis of atomic number}
\label{sec:AnZ}

\begin{figure}[b]\centering
\includegraphics[width=0.49\textwidth]{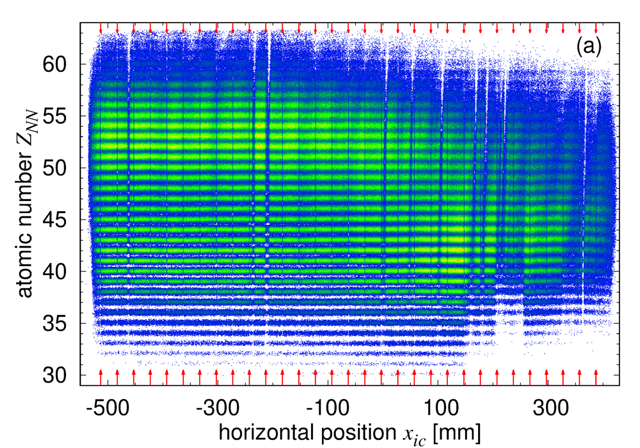}
\includegraphics[width=0.49\textwidth]{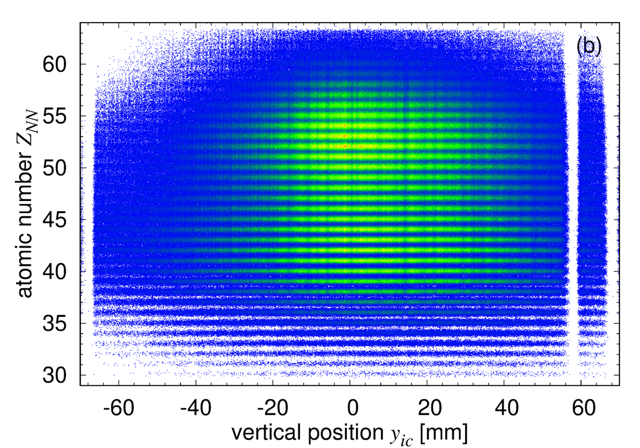}
\includegraphics[width=0.49\textwidth]{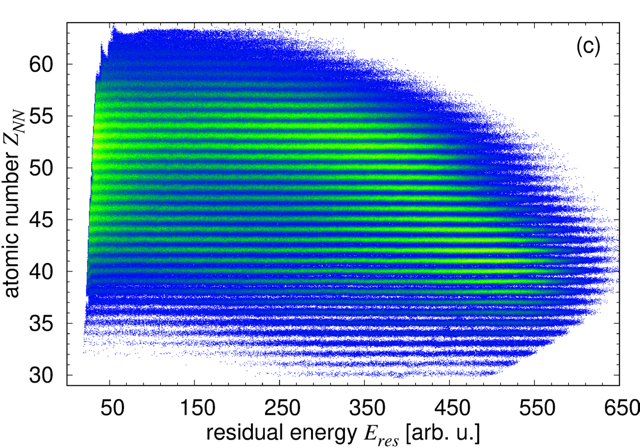}
\includegraphics[width=0.49\textwidth]{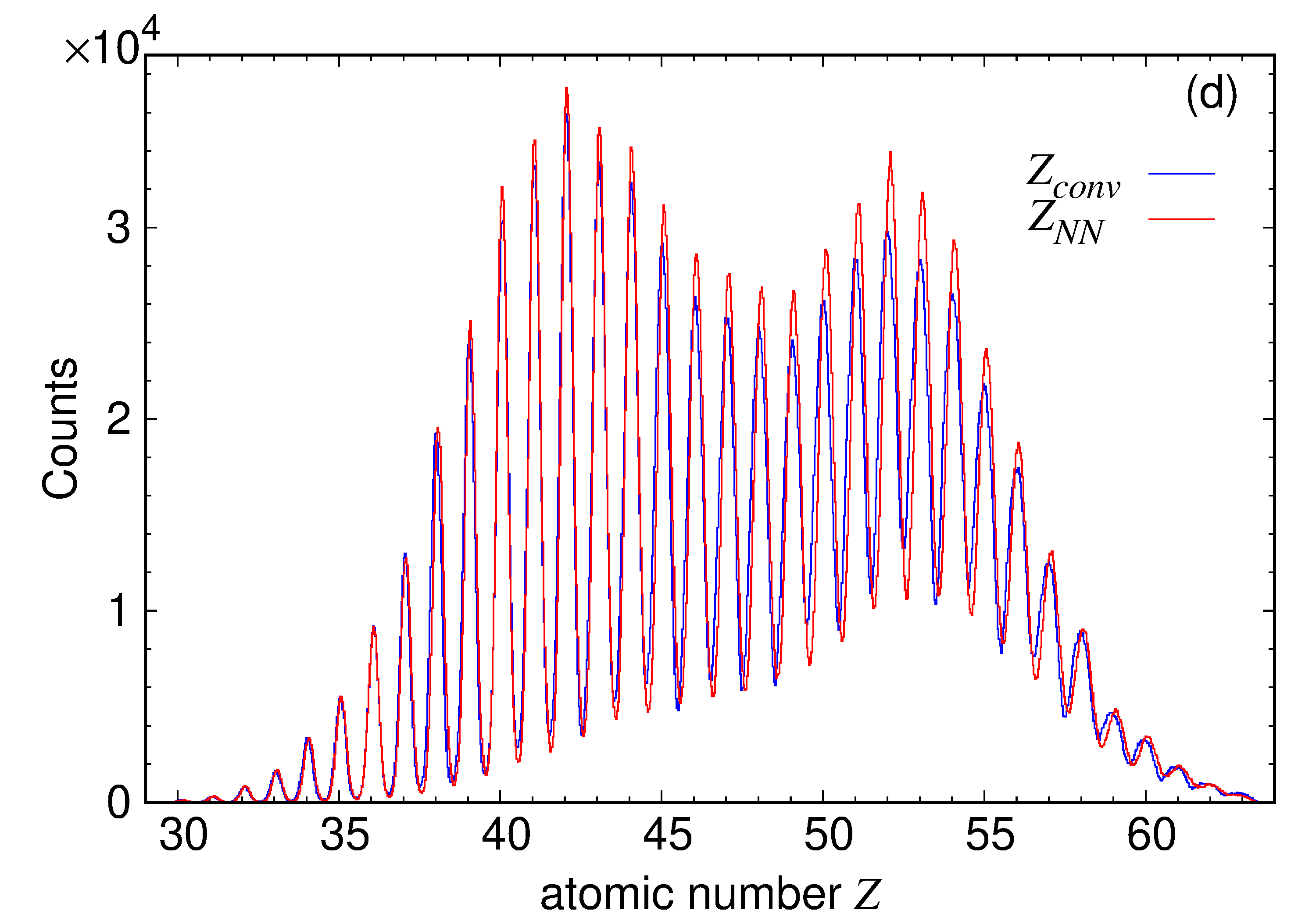}

\caption{Results of the neural network analysis of the atomic number: 
(a) and (b) The same as in Figure~\ref{fig:QXICYIC}(a) and (b) but for $Z_{NN}$ instead of $q$,
(c) Two-dimensional energy correlation plot between the atomic number ($Z_{NN}$) and residual energy ($E_{res}$)
resulting in correlated bands for each atomic number.
(d) One-dimensional spectrum of atomic number, in red $Z_{NN}$ and in blue $Z_{conv}$ obtained using the conventional
analysis method.
\label{fig:ZNNE}}
\end{figure}

To analyze the atomic number ($Z$) analogously one starts with the approximate calibration coefficients ($C_i$).
The resulting correlation between the energy loss ($\Delta E$) and residual energy ($E_{res}$) is depicted in 
Figure~\ref{fig:QMQdEE}(b). The different correlated bands correspond to different atomic numbers.
We have labeled the atomic numbers within the range $Z_l = [30,40]$, which corresponds to 
the repetition probability of approximately $13~\%$. The remaining portion of the data in the range $Z_u = [41,63]$ 
remained unlabeled. This selection is illustrated in Figure~\ref{fig:QMQdEE}(b) using red arrows. 
During the training of the neural network the gradient descent and the convergence were evaluated using 
the same approach as for the analysis of the atomic charge state. 
For the labeled events within the range $Z_l=[30,40]$, the training target was the corresponding $Z_t=Z_l$. 
Conversely, for the unlabeled events within the range $Z_u=[41,63]$, the target was the nearest integer value 
to the predicted value in each learning step $Z_p$, i.e., $Z_t=\lfloor Z_{p}+0.5 \rfloor$.

The input to the deep neural network comprised the following 15 variables:
\begin{itemize}
\item the raw energies ($E_{ic_i}$) with $i=0 \dots 9$,
\item the horizontal and vertical positions ($\delta x_{ic}$ 
and $x^{cnt}_{ic}$ defined above) and ($y_{ic}$),
\item the horizontal and vertical positions ($x_{mw}$) and ($y_{mw})$.
\end{itemize}

The training procedure applied was analogous to that described in Section~\ref{sec:AnQ}.
The corresponding training convergence curve of the RMSD for $Z$ as a function of time is depicted in Figure~\ref{fig:Conv},
The neural network achieved convergence, resulting in an RMSD for $Z$ of $0.227$.

The correlations between the atomic number ($Z_{NN}$) obtained by the neural network and the horizontal
and vertical position on the entrance of the ionization chamber are shown in Figures~\ref{fig:ZNNE}(a) and \ref{fig:ZNNE}(b), respectively.
Figure demonstrates a sharp definition of the $Z_{NN}$ lines and the absence of the dependencies on
the positions.
The correlation between the atomic number ($Z_{NN}$) and the previously defined residual energy ($E_{res}$) 
(equation~\ref{E}) is depicted in Figure~\ref{fig:ZNNE}(c).
As evident from the figure, the sharply defined correlated $Z$ bands are also independent of $E_{res}$.
The corresponding one-dimensional spectrum of $Z_{NN}$ is presented in Figure~\ref{fig:ZNNE}(d)
and compared to $Z_{conv}$ obtained in a complete process of the conventional analysis method.
The resulting resolution in terms of $\text{FWHM}(Z_{NN})/Z_{NN}$ 
ranges from $1.1~\%$, for the lowest, to $1.2~\%$, for the highest atomic number, respectively.
The overall improvement in terms of 
$\text{FWHM}$ for $Z_{NN}$ with respect to $Z_{conv}$ is $9~\%$.

\subsection{Isotopic identification}
\label{sec:II}

\begin{figure}[b]\centering
\includegraphics[width=0.49\textwidth]{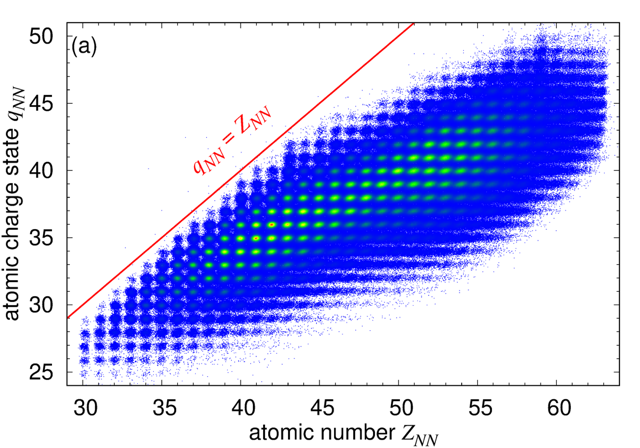}
\includegraphics[width=0.49\textwidth]{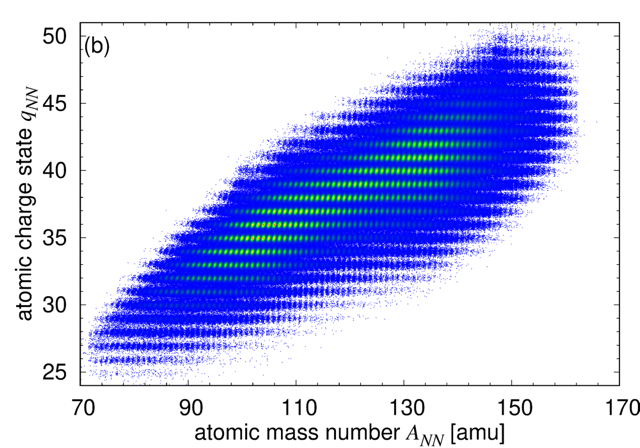}
\caption{Isotopic identification charts: (a) Two-dimensional correlation plot between the atomic charge state ($q_{NN}$) and 
atomic number ($Z_{NN}$). The red line corresponds to $q_{NN} = Z_{NN}$ for the fully stripped ions. 
(b) Two-dimensional correlation plot between the atomic charge state ($q_{NN}$) and atomic mass number ($A_{NN}$).
\label{fig:ZNNQNN}}
\end{figure}

The aforementioned results can now be utilized to complete the isotopic identification of ions detected by VAMOS++.
The two-dimensional correlation of the atomic charge state ($q_{NN}$) and the atomic number ($Z_{NN}$) is depicted 
in Figure~\ref{fig:ZNNQNN}(a). 
Figure~\ref{fig:ZNNQNN}(a) reveals the substantial width of the atomic charge state distributions for each element, 
with intensity gradually diminishing before reaching the fully stripped charge state, $q_{NN} = Z_{NN}$ (indicated 
by the red line). Notably, the peak of the $q_{NN}$ distribution is shifted by several units towards lower values. This shift is attributed to the nuclear reactions being induced at energies near the Coulomb barrier.
A small fraction ($2\times 10^{-6}$) of unphysical erroneous events can be observed beyond the $q_{NN} = Z_{NN}$.
These events are obtained using both conventional and neural network-based approaches.
The atomic mass number can be determined by combining the atomic charge state ($q_{NN}$) and 
the mass-over-charge, ($A/q$).
The correlation between the atomic charge state ($q_{NN}$) and the atomic mass number ($A_{NN}$) is depicted in 
Figure~\ref{fig:ZNNQNN}(b).

\section{Summary}
\label{sec:Summary}

In conclusion, precise analysis of the atomic charge state and atomic number is paramount for comprehensive isotopic identification of ions within the VAMOS++ spectrometer. This analysis is fundamentally linked to energy measurements 
from the segmented ionization chamber, yet it has been severely hampered by the unpredictable thickness and 
non-uniformity of the chamber's large-area entrance windows, along with other detector imperfections like 
charge collection issues and electronic non-linearities. Traditionally, achieving high-resolution atomic charge state 
and number through meticulous multivariate analysis has been an exceptionally lengthy and laborious process, 
often extending over several months.

This work successfully introduces a novel, highly efficient method for atomic charge state and number analysis, 
leveraging the power of deep neural networks. These networks are uniquely capable of discerning intricate patterns 
and generalizing them to previously unseen data. Crucially, the networks were trained on an experimental dataset 
with only a minimal fraction of precisely labeled events (corresponding to the lowest and best-resolved states/numbers), enabling them to autonomously and accurately process the vast majority of unlabeled events. This innovative approach dramatically reduces the analysis time, allowing for the rapid acquisition of high-resolution atomic charge state and 
number spectra in mere hours.
By discarding the human factor, our method ensures the efficient acquisition of standardized, optimal, and highly 
reproducible results. This breakthrough has yielded exceptional resolutions: $\text{FWHM}(q_{NN})/q_{NN}$ ranges 
from $1.2~\%$ to $1.0~\%$ for atomic charge states $q_{NN} = [25, 50]$, and $\text{FWHM}(Z_{NN})/Z_{NN}$ ranges from $1.1~\%$ to $1.2~\%$ for atomic numbers $Z_{NN} = [30, 63]$. These results represent a significant advancement in the capabilities of VAMOS++ data analysis.

\providecommand{\href}[2]{#2}\begingroup\raggedright\endgroup

\end{document}